\documentclass[%
 aip,
 amsmath,amssymb,
 reprint,%
]{revtex4-1}

\usepackage{graphicx}
\usepackage{dcolumn}
\usepackage{bm}

\usepackage[utf8]{inputenc}
\usepackage[T1]{fontenc}
\usepackage{mathptmx}
\usepackage{etoolbox}
\usepackage{bigints}
\usepackage[dvipsnames]{xcolor}
\usepackage{comment} 
\usepackage{multirow}

\makeatletter
\def\@email#1#2{%
 \endgroup
 \patchcmd{\titleblock@produce}
  {\frontmatter@RRAPformat}
  {\frontmatter@RRAPformat{\produce@RRAP{*#1\href{mailto:#2}{#2}}}\frontmatter@RRAPformat}
  {}{}
}%
\makeatother
\begin{document}


\title{On the interfacial properties of hydroquinone: realistic and coarse-grained molecular models from computer simulation}

\author{M. J. Torrejón}
\affiliation{Laboratorio de Simulaci\'on Molecular y Qu\'imica Computacional, CIQSO-Centro de Investigaci\'on en Qu\'imica Sostenible and Departamento de Ciencias Integradas, Universidad de Huelva, 21006 Huelva Spain}

\author{B. Rodríguez García}
\affiliation{Departamento de Física Aplicada, Universidade de Vigo, 36310, Spain}

\author{J. Algaba}
\affiliation{Laboratorio de Simulaci\'on Molecular y Qu\'imica Computacional, CIQSO-Centro de Investigaci\'on en Qu\'imica Sostenible and Departamento de Ciencias Integradas, Universidad de Huelva, 21006 Huelva Spain}

\author{J. M. Olmos}
\affiliation{Laboratorio de Simulaci\'on Molecular y Qu\'imica Computacional, CIQSO-Centro de Investigaci\'on en Qu\'imica Sostenible and Departamento de Ciencias Integradas, Universidad de Huelva, 21006 Huelva Spain}

\author{M. Peréz-Rodríguez}
\affiliation{Instituto de Química Física Blas Cabrera, CSIC, Madrid, E-28006, Spain}

\author{J. M. Míguez}
\affiliation{Laboratorio de Simulaci\'on Molecular y Qu\'imica Computacional, CIQSO-Centro de Investigaci\'on en Qu\'imica Sostenible and Departamento de Ciencias Integradas, Universidad de Huelva, 21006 Huelva Spain}

\author{A. Mejía}
\affiliation{Departamento de Ingeniería Química, Universidad de Concepción, POB 160-C, Concepción, Correo 3, Chile}

\author{M. M. Piñeiro}
\affiliation{Departamento de Física Aplicada, Universidade de Vigo, 36310, Spain}

\author{F. J. Blas$^{*}$}
\affiliation{Laboratorio de Simulaci\'on Molecular y Qu\'imica Computacional, CIQSO-Centro de Investigaci\'on en Qu\'imica Sostenible and Departamento de Ciencias Integradas, Universidad de Huelva, 21006 Huelva Spain}

\begin{abstract}
In this work, we determine the vapor-liquid (VL) coexistence and interfacial properties of the hydroquinone (HQ) pure system from $NVT$ molecular dynamics simulations. We employ the direct coexistence technique to put in contact both phases in the same simulation box and generate the VL interface. Five different models have been tested to describe the HQ molecule in order to assess the performance of different approaches. The first two models are based on the Transferable Parameters Potentials for Phase Equilibria (TraPPE) force field. The first TraPPE model is the original one [\emph{J. Chem. Phys. B} \textbf{111}, 10790–10799 (2007) and \emph{J. Chem. Phys. B} \textbf{117}, 73–288
(2013)] based on an all-atoms approach (TraPPE-AA). The second TraPPE model is proposed for the first time in this work and is based on an united-atoms approach (TraPPE-UA) where the --CH groups from the aromatic ring are modeled as a single interaction site. We also use two HQ models based on the Optimized Potentials for Liquid Simulations (OPLS) force fields. Both OPLS models have already been reported in the literature [\emph{J. Am. Chem. Soc.} \textbf{118}, 11225–11236 (1996) and \emph{J. Chem. Phys.} \textbf{148}, 244502 (2018)], but this is the first time that are used to describe VLE and interfacial behavior. In addition, we propose a new Coarse grain (CG) HQ model based on the Statistical Associating Fluid Theory (SAFT) framework. We determine density profiles, coexistence densities, vapor pressure, interfacial thicknesses, and interfacial tensions as obtained from the $NVT$ simulations with the five different models. We explore the VL behavior of pure HQ system from 500 to $750\,\text{K}$. Remarkably good agreement has been found between the simulation results obtained by the TraPPE-AA, CG, and both OPLS models. Unfortunately, the results obtained by the TraPPE-UA model proposed in this work show discrepancies with the rest of the HQ models. Finally, we also determine the critical temperature, density, and pressure from the analysis of the coexistence densities, vapor pressure, and interfacial tension. The critical temperature predicted by the TraPPE-AA is in excellent agreement with the experimental data taken from the literature. The CG and both OPLS HQ models slightly underestimate experimental data, and the TraPPE-UA model clearly overestimates it.
\end{abstract}

\maketitle
$^*$Corresponding author: felipe@uhu.es




\section{Introduction}

Hydroquinone (also known as 1,4-dihydroxybenzene, quinol, or HQ hereafter for brevity) is among the ensemble of organic molecules able to form clathrates. In this particular case, the crystalline inclusion solid formed by HQ as host molecule has attracted interest in the last decades due to some particular properties. Daschbach \emph{et al.}~\cite{Daschbach2006} first introduced an inspiring Molecular Simulation study that postulated a high H$_2$ storage capacity for this system, analyzing the multiple guest occupation modes within the crystalline cells in the clathrate channels. In a series of very interesting experimental articles, the research group of Prof.~Yoon demonstrated that HQ clathrate presents an enhanced affinity to capture CO$_2$ as guest molecule, with remarkable selectivity towards other gases in a mixture~\cite{Lee2011a, Lee2011b,Lee20167604}. In addition, the same group evidenced again that this clathrate is a potential candidate to store H$_2$ in an accessible and reversible way~\cite{Han2012120}. These two singular features alone suffice to make HQ molecule an extremely appealing research objective due to the potential applications involved. These initial studies promoted further analysis on experimental HQ clathrate characterization~\cite{Torre20165330,Coupan201735,Torre201914582,Rozsa20141880}. From a theoretical perspective, Conde \emph{et al.} \cite{Conde201610018} studied the HQ clathrate thermodynamics applying the well known classical approximation of van der Waals and Plateeuw~\cite{Platteeuw1957a,Platteeuw1959a}, widely used to study the phase equilibria of hydrates and clathrates. Concerning the mentioned potential applications of HQ clathrate, the possibility of finding an alternative to current H$_2$ storage techniques justifies further analysis and studies towards a better understanding of these clathrate properties. The recent review of Gupta \emph{et al.}~\cite{Gupta202169} describes in detail the potential of hydrates and organic clathrates as the forthcoming generation of large scale hydrogen storage materials, citing among other options HQ as a promising candidate.

A number of recent studies have been also devoted to the estimation of structural, thermodynamic and diffusion properties of HQ clathrates using different Molecular Simulation techniques~\cite{Comesana2018a,Perez-Rodriguez201818771,Rodriguez-Garcia2023,Parage20246939}. These studies have brought some insight into the properties of HQ clathrates, opening promising perspectives. Despite this, its complete characterization is still very far from being satisfactory. For instance, the thermodynamic properties of fluid phases of HQ are still poorly known from both experimental and theoretical points of view, and their precise characterization is crucial to ascertain the feasibility of these clathrates in the practical applications noted.

Considering the case of HQ clathrates, and from a molecular simulation perspective, these studies require first a careful and quantitative evaluation of the performance of the existing HQ molecular models in the estimation of its phase equilibria and termophysical properties. In this context, the present article is a compulsory step, analyzing the estimation of HQ VLE using different molecular modeling approaches, leading to a discussion about the optimal representation approach required towards further analysis of the feasibility limits of this alternative to other existing H$_2$ storage materials. 

Although it would be interesting to validate the results obtained from molecular dynamics simulations with experimental results reported in the literature in the whole range of thermodynamic conditions studied in this work, there is a clear lack of experimental data reported in the literature. As far as the authors know, there is very limited experimental data on vapor-liquid phase equilibria and critical properties in the literature. From the NIST database~\cite{NIST_AM} it is possible to collect the vapor and liquid densities along the whole VL equilibria curve as well as the critical temperature and critical density point. However, the vapor pressure is only reported at low temperatures up to $600\,\text{K}$ in the NIST database. Concerning the critical temperature and pressure, only two experimental measures have been found in the literature by Liessmann \emph{et al.}~\cite{Liessmann1995a} and  Gmehling \emph{et al.}~\cite{Gmehling1983a} reported a similar critical temperature value (820 and $818\,\text{K}$, respectively) but different critical vapor pressures (67.5 and $61.8\,\text{bar}$, respectively). Due to the lack of experimental data, and the differences between the experimental data values reported, it is complicated to provide an accurate description of the vapor-liquid equilibria, and the critical and interfacial properties of the HQ system. This work aims to improve the knowledge of the HQ pure system and shall be considered as the first step toward the study of more complex systems as HQ clathrates.

The organization of this paper is as follows. In Section 2, we describe the molecular models used in this work. Section 3 is devoted to the simulation details. The results obtained, as well as their discussion, are described in Section 4. Finally, conclusions are presented in Section 5.

\section{Molecular models}

In this work, five different models have been used to describe hydroquinone (HQ) molecule. The first two models are based on the widely-known TraPPE (Transferable Potentials for Phase Equilibria) force field parameters,~\cite{Rai2007a,Rai2013a,Wick2000a} the next two HQ models are based on the OPLS force fields\cite{Jorgensen1993a,Jorgensen1996a} (Optimized Potentials for Liquid Simulations), and finally, a Coarse-Grained (CG) model based on the Mie potential is used.~\cite{Mie1903a,Muller2014a,Muller2017a} Three of the models used in this work (TraPPE-AA and both OPLS models) have been taken from literature. However, this is the first time that these models are used to describe vapor-liquid (VL) phase equilibria and interfacial properties of the pure HQ molecule. Also, two new models are proposed in this work. One of them is based on the TraPPE force field parameters, and the other one is based on the CG approach and has been developed using the SAFT (Statistical Associating Fluid Theory) theoretical framework. In this section, we provide a detailed analysis of the five models. We have also included a detailed description of the molecular models employed in this work as a Supplementary Material. In addition, readers can request access to the input files if needed and we will provide the necessary information.

\subsection{TraPPE models}

Two different TraPPE HQ models have been used in this work. The difference between them is that two different approaches have been used to describe the H atoms within the aromatic ring. In the first one, H atoms are explicitly taken into account by an all-atoms model (TraPPE-AA),~\cite{Rai2007a,Rai2013a}  while in the second one, the aromatic CH group is considered as a unique interaction center by a united-atoms model (TraPPE-Hybrid-UA). It is important to remark that the TraPPE-Hybrid-UA HQ model is proposed for the first time in this work. This new TraPPE-Hybrid-UA model can be considered as the combination of the original TraPPE-AA HQ model~\cite{Rai2007a,Rai2013a} and the original TraPPE-UA benzene model~\cite{Wick2000a}. It is important to note that the TraPPE-Hybrid-UA model proposed here is not a purely TraPPE-UA force field in the strictest sense, as it combines both AU and AA atoms. A true TraPPE-UA model would exclusively use UA parameters to consistently represent hydroquinone. Because of this, we use the name TraPPE-Hybrid-UA instead of TraPPE-UA force field model. From the TraPPE-UA benzene model, the non-bonded CH group parameters and the CH-CH and CH-C distances were taken. The rest of the necessary parameters for describing the HQ molecule were taken from the TraPPE-AA HQ model (including non-bonding and bonding interactions). It is also important to remark that following the TraPPE-UA approach, the H atoms from the hydroxyl (OH) groups are still explicitly considered. 

In both TraPPE models, the force field parameters for the non-bonded interactions are described by the Lennard-Jones (LJ) and Coulomb potentials for the dispersive and coulombic interactions:

\begin{equation}
U(r_{ij})=4\epsilon_{ij}\left[\left(\frac{\sigma_{ij}}{r_{ij}}\right)^{12} -
\left(\frac{\sigma_{ij}}{r_{ij}}\right)^{6}\right] + \frac{q_iq_j}{4\pi\epsilon_0r_{ij}}
\label{EQ:non-bonded}
\end{equation}

\noindent
where $r_{ij}$ is the $i$ and $j$ chemical group distance, $\sigma_{ij}$ and $\epsilon_{ij}$ are the diameter and well depth associated with the LJ potential, $q_i$ and $q_j$ are the partial charges placed on chemical groups $i$ and $j$, and $\epsilon_0$ is the permittivity of vacuum. The TraPPE-AA molecular parameters of the HQ molecule can be found in the original work of Rai and Siepmann.~\cite{Rai2007a,Rai2013a} As pointed out previously, the TraPPE-Hybrid-UA molecular parameters are taken from the original TraPPE-AA HQ model except for the non-bonded CH group parameters and the CH-CH and CH-C distance that are taken from the TraPPE-UA benzene model.

TraPPE models are characterized to be semiflexible models where the distance between interaction centers is fixed, but those interaction centers separated by two bonds interact through a harmonic bending potential and those separated by three bonds interact through a torsional potential. However, due to the scarce flexibility of the aromatic ring, in both TraPPE models, the aromatic ring is modeled as planar and rigid, i.e. bending and torsional degrees of freedom are fixed to the molecular equilibrium values. However, there is still some bending and torsional contribution through the hydroxyl (-OH) groups. For further details, we refer the reader to the original works of Rai and Siepmann~\cite{Rai2007a,Rai2013a} since the bending and torsional parameters are the same in both TraPPE models and equal to the original TraPPE-AA HQ model.

\subsection{OPLS models}

In this work, two OPLS models for describing the HQ molecules have been used. In both cases, the H atoms from the aromatic ring and the hydroxyl groups are taken explicitly into account, i.e., both OPLS models are based on an all-atoms approach (OPLS-AA). The first OPLS model is the original one proposed by Jorgensen \emph{et al.}~\cite{Jorgensen1993a,Jorgensen1996a} (named simply OPLS from now on). This model is fully-flexible, where the atoms separated by one bond interact through a bonding harmonic potential, atoms separated by two bonds interact through a bending harmonic potential, and atoms separated by three bonds interact through a torsional potential. The force field parameters for describing internal degrees of freedom of both OPLS HQ models used in this work can be found in the original works of Jorgensen \emph{et al.}~\cite{Jorgensen1993a,Jorgensen1996a}

In both OPLS models, non-bonded interactions are described by Eq.~\eqref{EQ:non-bonded}. The parameters used for describing the LJ interactions in both models are the same as those proposed by Jorgensen \emph{et al.}~\cite{Jorgensen1993a,Jorgensen1996a} However, some of us proposed in a previous work a reparametrization of the partial charges located in each atom of the HQ molecule.~\cite{Comesana2018a} In this new OPLS model (called OPLS-MOD from now on), the partial charges located at each atom have been recalculated using charges from electrostatic potentials using a grid-based method scheme.~\cite{Breneman1990a} The optimized partial charges for the OPLS-MOD HQ model have been described in detail in our previous work.~\cite{Comesana2018a} This molecular model has been used to describe the structure and properties of different HQ clathrate structures\cite{Perez-Rodriguez201818771,Torre201914582,Mendez-Morales2022,Rodriguez-Garcia2023,Parage20246939}.

\subsection{CG model}

The methodology used to achieve this force field is based on a top-down approach, which contrasts with the more conventional bottom-up methodology. The latter relies on a quantum-mechanical or atomic-level description of molecular interactions, incorporating certain degrees of freedom from the real system through various techniques and refining model parameters via iterative simulations. This philosophy has been the foundation of well-established all-atom and united-atom force fields since the early 1980s.

\begin{figure}
\begin{centering}
\includegraphics[scale=0.3]{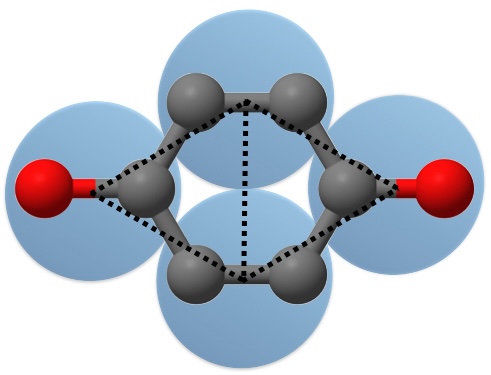}
\par\end{centering}
\caption{\label{fig:CG}Schematic representation of hydroquinone using a four-site coarse-grained model. Grey spheres correspond to the aromatic carbon groups (CH and C), red spheres to the hydroxyl groups, and blue spheres to the four Mie potential interacting centers.}

\end{figure}

In contrast, the top-down methodology employed in our CG model for HQ is based on an accurate equation of state that links macroscopic fluid properties to force field parameters. Specifically, we utilize the Statistical Associating Fluid Theory for rings~\cite{Muller2014a,Muller2017a}, which has demonstrated remarkable accuracy in predicting phase equilibria of complex molecular systems, to model HQ as a planar molecule formed by four identical tangent spheres (interacting via the Mie potential~\cite{Mie1903a}),
bonded rigidly at a distance equal to their diameter from the center
of adjacent beads and forming two equilateral triangles, as it is
observed in Figure~\ref{fig:CG}. The main obvious advantage is that HQ is modeled as a ring formed from four monomeric units, a much less number of interaction sites compared with the AA models ($14$ interaction sites), which considerably reduces the CPU time needed to simulate with confidence the phase equilibria and interfacial properties of HQ and with a similar accuracy than most of the force fields based on the bottom-up methodology analyzed in this work. In addition to this, it enables direct simulations to determine a wide range of thermodynamic, structural, interfacial, and dynamic properties. Notably, this top-down strategy has not been previously applied to describe the thermodynamic or interfacial properties of HQ. Additionally, recent studies have highlighted the capability of CG molecular models to capture complex interfacial behavior in multicomponent mixtures, sparking valuable discussions about their quantitative performance compared to atomistic molecular models. However, the application of this approach to predict stable solid phases of HQ may be limited. A detailed representation of hydrogen-bonding interactions between the hydroxyl groups is crucial for accurately modeling HQ crystalline structures, which the CG model may not fully capture.


Although it is possible to describe heterogeneous molecules with a heterogeneous SAFT group contribution approach where each segment is described by different Mie parameters, the isotropic segment approach has been extensively used in the literature, showing great agreement with the experimental data~\cite{Muller2017a,Avendano2011a,Avendano2013a,Lafitte2012a,Mejia2014b,Cumicheo2014a,Lobanova2015a}.

In this molecular model, the spheres interact with each other according
to the Mie molecular potential~\cite{Mie1903a}:

\begin{equation}
\phi_{Mie}\left(r\right)=C\varepsilon\left[\left(\frac{\sigma}{r}\right)^{\lambda_{r}}-\left(\frac{\sigma}{r}\right)^{\lambda_{a}}\right]
\label{eq:Mie2}
\end{equation}

\noindent where $r$ is the distance between the interaction centers of the particles, $\sigma$ and $\epsilon$ are the diameter and well depth associated with the Mie potential, and $\lambda_{r}$ and $\lambda_{a}$ are the repulsive and attractive
exponents, respectively, that characterize the range of the interaction. $C$ is a constant given in terms of repulsive and attractive exponents,

\begin{equation}
C=\left(\frac{\lambda_{r}}{\lambda_{r}-\lambda_{a}}\right)\left(\frac{\lambda_{r}}{\lambda_{a}}\right)^{\lambda_{a}/\left(\lambda_{r}-\lambda_{a}\right)}
\label{eq:Mie}
\end{equation}

In this work, these values are obtained by invoking the corresponding
state principle described in a previous work,~\cite{Muller2017a} and their
numerical values are summarized in Table~\ref{tab:Mie-CG}.

\begin{table}
\caption{\label{tab:Mie-CG}Mie Molecular Parameters for hydroquinone.}

\centering{}%
\begin{tabular}{cccc}
\hline\hline
$\varepsilon/k_{B}$ (K) & $\sigma$ (\AA) & $\lambda_{r}$ & $\lambda_{a}$\tabularnewline
\hline
455.948 & 3.230 & 19.425 & 6\tabularnewline
\hline\hline
\end{tabular}
\end{table}

\section{Simulation details}

All molecular dynamics simulations have been carried out using the open-source software GROMACS (version 2016, double-precision) package. In order to study the VL equilibria predicted by the five different models, we perform simulations in the canonical or $NVT$ ensemble. We have used a Verlet leapfrog algorithm~\cite{Cuendet2007a}  with a time step of $0.001\,\text{ps}$ for the TraPPE and OPLS models and a time step of $0.003\,\text{ps}$ for the CG model. The temperature is fixed along the simulation by the Nosé -- Hoover thermostat\cite{Nose1984a} with
a relaxation constant of $1.0\,\text{ps}$. Also, we use periodic boundary conditions in all three directions. In all cases, we study the VL equilibria and interfacial properties at six different temperatures ($500$, $550$, $600$, $650$, $700$, and $750\,\text{K}$). In this work, the VL phase behavior is studied by placing both bulk phases, the liquid, and the vapor phases, in contact in the same simulation box. This initial setup configuration allows the system to evolve towards VL equilibrium conditions. This technique, known as direct coexistence technique, allows to study not only the VL equilibrium but also the interfacial properties. At this point is important to remark that the direct coexistence technique is not the most accurate method to determine vapor-liquid coexistence conditions. This is because the sampling of this technique produces larger deviations when dealing with low vapor pressures and densities~\cite{Vega2007a}. However, this technique is especially suitable for studying interfacial properties since the interface is directly generated in the simulation box, which allows the study of the interfacial properties directly from the simulation. However, if one is most interested in the equilibrium properties, especially in the vapor density and pressure at low temperatures, more rigorous techniques such as the Gibss ensemble should be employed~\cite{Cortes2013a}.

In particular, at each temperature, we estimate the equilibrium densities of the coexisting liquid and vapor phases, the vapor pressure, as well as the interfacial tension and the interfacial thickness. Finally, we also obtain the critical temperature, pressure, and density predicted by the five HQ models used.

The initial simulation setup for the OPLS and TraPPE models is built up in the same manner. First, an initial $L_x=L_y=5$ and $L_z=15\,\text{nm}$ parallelepipedic  box is filled with 2000 HQ molecules. After an energy minimization stage, the $L_z$ side of the simulation box is elongated from 15 to $40.5\,\text{nm}$, and the HQ molecules are displaced to the center of the new simulation box. With this procedure, we create a box with the HQ molecules placed in the center of the box (liquid phase), surrounded by two empty spaces in the $z$-axis direction. Once the simulation starts, HQ molecules from the liquid phase may jump to the initial empty space to generate two vapor phases at each side of the centered liquid phase, \textit{i.e.}, two VL interfaces are created. The final dimension box, as well as the number of molecules placed in it, is large enough to ensure that both phases in equilibria can be developed in the same simulation box. In these cases, a cutoff value of $1.1085\,\text{nm}$ is used to truncate dispersive and coulombic interactions. This cutoff value corresponds to 3 times the largest $\sigma$ value of the four models (TraPPE and OPLS). In particular, $\sigma$ corresponds to that used to describe the CH group in the TraPPE-Hybrid-UA model. We use particle mesh Ewald (PME) \cite{Essmann1995a,Lundberg2016a} to deal with long-range corrections (LRCs) for both dispersive and coulombic interactions. The advantage of the PME LRCs corrections, over the classical one,~\cite{Allen1987a,Frenkel2002a} is that PME LRCs can take into account the inhomogeneity of the system and can be applied when two or more phases are present in the same simulation box. It is important to notice that a cut-off value above six times the diameter of the biggest molecule group provides a reliable description of the interfacial properties when particles interact through the Lennard-Jones potential~\cite{Allen1987a,Frenkel2002a,Muller2021a,Galliero2010a,Miguez2013a,Martinez-Ruiz2014a}. In this work, we use a fixed cut-off value that corresponds to 3 times the largest $\sigma$ value of the four LJ models (CH TraPPE-Hybrid-UA group) and, additionally, PME LRCs are applied to analytically estimate the LJ interactions beyond the cut-off value. It is important to take into account that the use of a short cut-off value + LRCs provides the same results as a large cut-off value. Finally, simulations are run for $25\,\text{ns}$, taking only the last $20\,\text{ns}$ as production period.

For the CG simulations, we follow a different procedure.\cite{Muller2021a} We place 2115 HQ molecules (8640 beads) in a simulation box in which the total volume, $V$, is calculated using the bulk density average, $\left(\rho^{L}+\rho^{V}\right)/2$,
at the lowest temperature ($500\,\text{K})$. The numerical values of $\rho^{L}$,
and $\rho^{V}$ are taken from NIST database.\cite{NIST_AM} The
simulation volume, $L_{x}\times L_{y}\times L_{z}$, is contained in a  parallelepipedic box. We use periodic boundary conditions in all three directions. Again, $L_{i}$
values are chosen in order to have a cell large enough to accommodate
liquid and vapor regions with enough molecules to ensure a sensible
amount of bulk of both phases. The numerical values of the simulation
box are $L_{x}=L_{y}=41\textrm{Å}$, and $L_{z}=10\times L_{x}=410\textrm{Å}$.
In order to speed up the formation and stabilization of the bulk phase
and interfacial region, the system is initially held at a high temperature,
above its critical state $\left(T_{c}=828.3\,\text{K}\right)$, where a unique
homogeneous well-mixed phase is present. The system is then quenched
to the desired temperature allowing it to evolve under $NVT$ conditions
until equilibration is reached through diffusive mass transport. To
reduce the truncation and system size effects involved in the phase
equilibrium and interfacial properties calculations, a cutoff radius
of $19.41\,\text{\AA}$ $\left(r_{c}\approx6\sigma\right)$ is
used throughout. Notice that in GROMACS it is not possible to apply PME long-range corrections when a different potential than LJ is applied. However, a cutoff value of $6\,\sigma$ is enough to accurately describe the VL equilibria and interfacial properties~\cite{Galliero2009a,Galliero2010a,Miguez2013a,Martinez-Ruiz2014a}. After the initial temperature quenching,
the systems are equilibrated for 30 ns, and then a production run
is set for at least another 60 ns.

In this work, since we have a planar VL interface, the equilibrium vapor pressure, $P$, corresponds to the normal component of the pressure tensor, $P_{zz}$. Also, from the diagonal components of the pressure tensor, it is possible to calculate the interfacial tension following the mechanical route~\cite{Hulshof1901a,deMiguel2006a,deMiguel2006b}:

\begin{equation}
\gamma=\frac{L_{z}}{2}\left[P_{zz}\left(z\right)-\frac{P_{xx}\left(z\right)+P_{yy}\left(z\right)}{2}\right]
\label{iftmd}
\end{equation}

\noindent where $P_{kk}(z)$ are the pressure tensor components (being $k$  either $x$, $y$ or $z$) as a function of the coordinate perpendicular to the interface, $z$, and $L_z$ is the length of the simulation box along the $z$ direction. Finally, the $1/2$ factor arises to take into account that the simulated system presents two VL interfaces.

In this work, we also estimate the critical temperature, $T_c$, density, $\rho_c$, and pressure, $P_c$, from the VL coexistence equilibrium results by the use of the scaling and the rectilinear diameters laws, respectively~\cite{Rowlinson1982a,Xiang2005a}:

\begin{equation}
\rho_{L}-\rho_{V} = A\left(T-T_{c}\right)^{\beta}
\label{tcscaling}
\end{equation}

\begin{equation}
\frac{\rho_{L}+\rho_{V}}{2} = \rho_{c}+B\left(T-T_{c}\right)
\label{rhocscaling}
\end{equation}

\noindent
Here $\beta$ is the corresponding critical exponent which has a universal value of $\beta=0.325$,\cite{Rowlinson1982b} and
$A$, $B$, $T_{c}$
and $\rho_{c}$ are obtained by 
fitting Eqs.~(\ref{tcscaling}) and (\ref{rhocscaling}) to the VL equilibria simulation results. $\rho_{L}$ and $\rho_{V}$ are the liquid and vapor coexistence
densities at the corresponding temperature $T$ obtained from the different HQ models. Also, it is possible to calculate $T_c$ following a similar approach as in Eqs.~(\ref{tcscaling}) and (\ref{rhocscaling}) but using the interfacial tension, $\gamma$, simulation data instead of the equilibrium densities. We can apply the scaling laws using the interfacial tension for fitting the following expression~\cite{Widom1965a}:

\begin{equation}
\gamma=\gamma_{0}\left(1-T/T_{c}\right)^{\mu}
\label{gammacscalng}
\end{equation}

\noindent
where $\gamma_{0}$ is the so-called "zero-temperature" surface tension
and $\mu$ is the corresponding critical exponent which has a universal value of
$\mu=1.258$~\cite{Rowlinson1982b}. Once again, the
unknown constants, $\gamma_{0}$ and $T_{c}$ are found by fitting Eq.~(\ref{gammacscalng}) to the interfacial tension simulation data obtained in this work. Finally, the critical pressure is estimated from
an extrapolation of the Clausius-Clapeyron relation to the critical temperature. Notice that the critical temperature has been already obtained
from Eqs.~(\ref{tcscaling}) and (\ref{gammacscalng}):

\begin{equation}
\ln P=C_{1}+\frac{C_{2}}{T}
\label{pcscaling}
\end{equation}

\noindent
where $C_{1}$, and  $C_{2}$ are obtained by fitting Eq.~\eqref{pcscaling} to the values of pressure obtained in this work at each temperature. The critical properties estimated in this work from the VL equilibria simulation results in combination with Eqs.~\eqref{tcscaling}-\eqref{pcscaling} are presented in Table \ref{table3}.

\section{Results}

In this section, we focus on the analysis of the VL coexistence equilibria and interfacial properties of the HQ system predicted by the five different models employed in this work. In particular, we obtain the density profiles, coexistence densities, vapor pressures, interfacial thicknesses, and interfacial tensions. We also analyze the behavior of these properties as a function of the temperature and we estimate the critical temperature, pressure, and density predicted by the five HQ molecular models. Finally, we compare our results with experimental data taken from the literature when it is available.

\subsection{Density profiles}

\begin{figure}
\centering
\includegraphics[width=0.8\columnwidth]{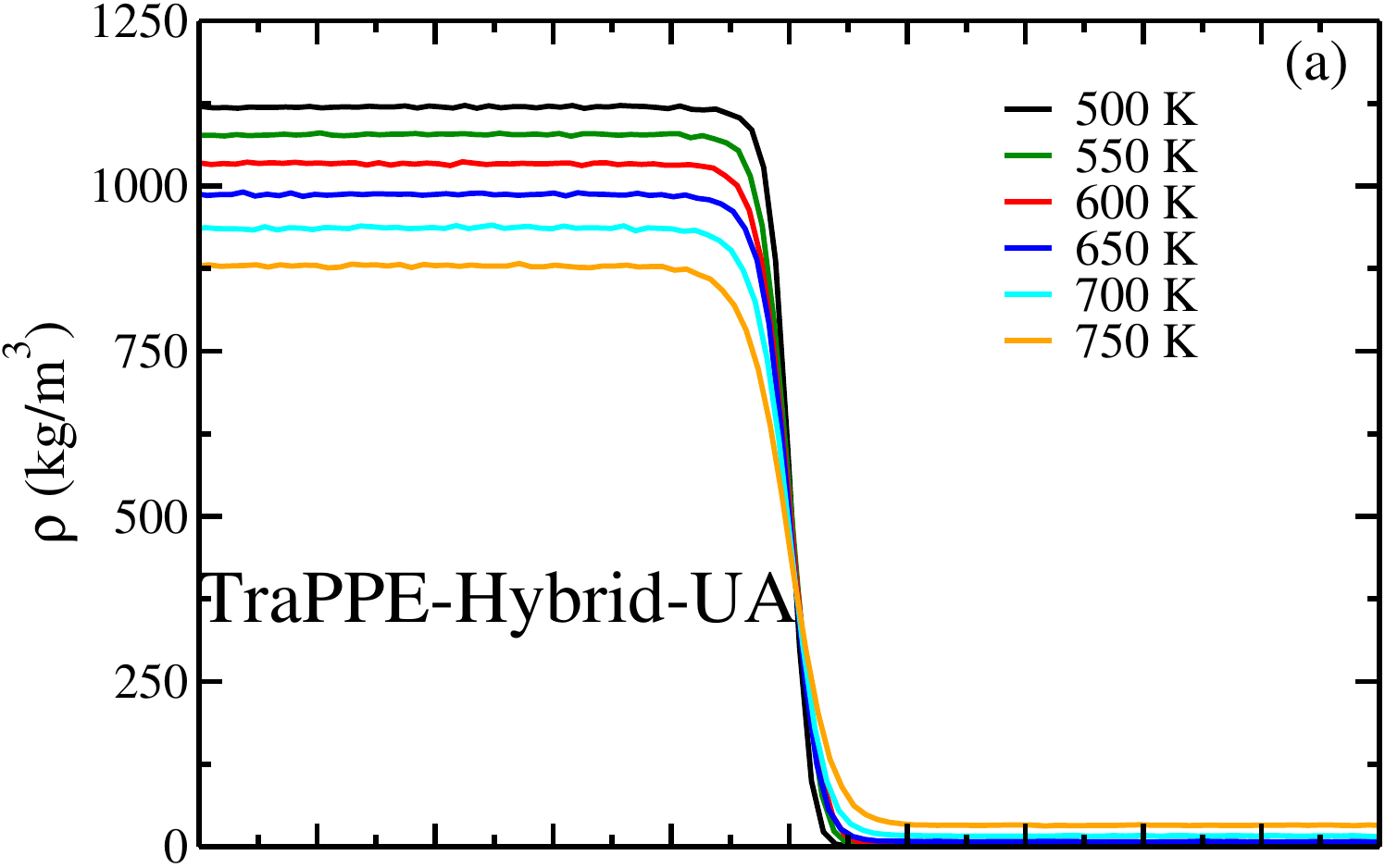}
\includegraphics[width=0.8\columnwidth]{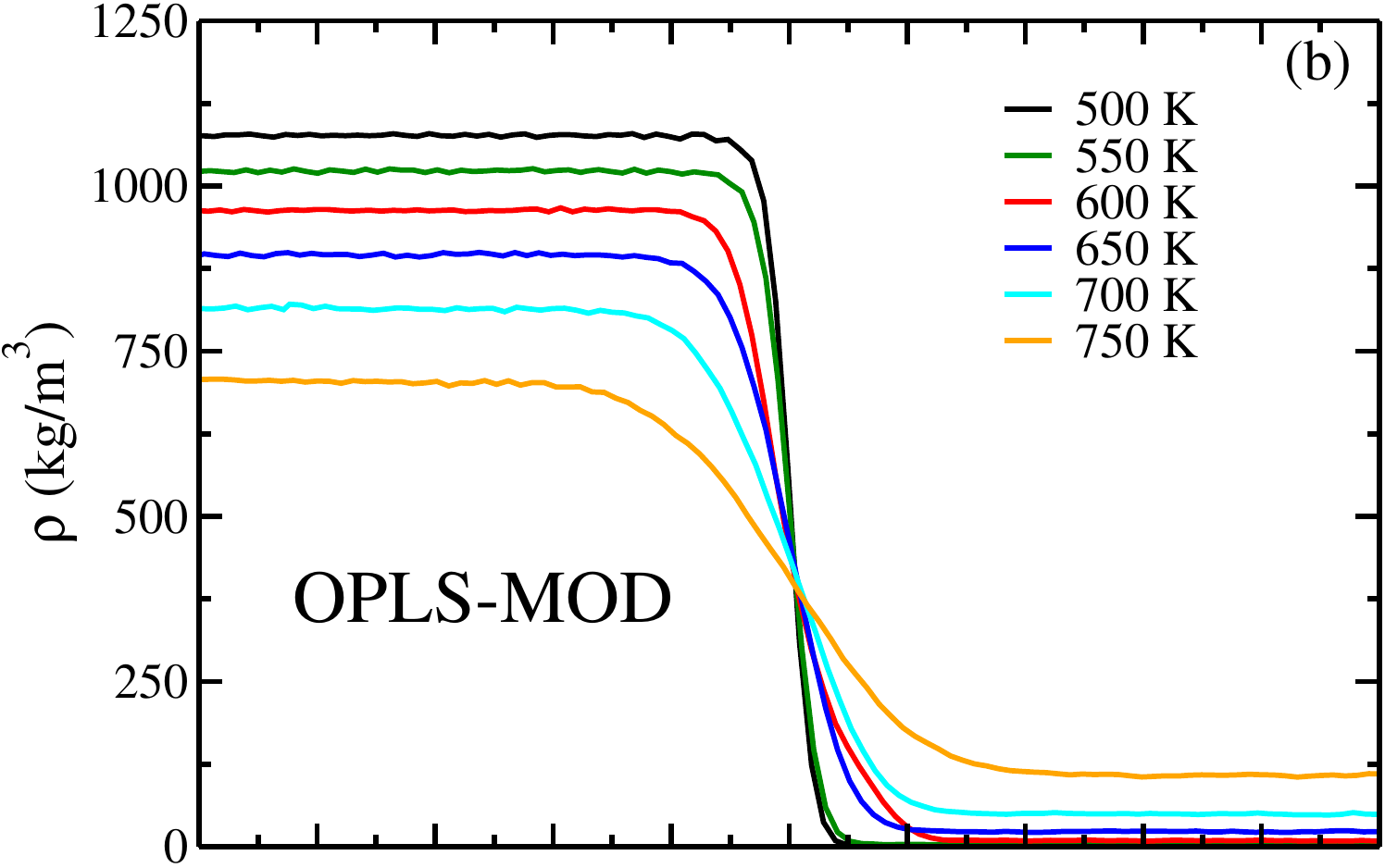}
\includegraphics[width=0.82\columnwidth]{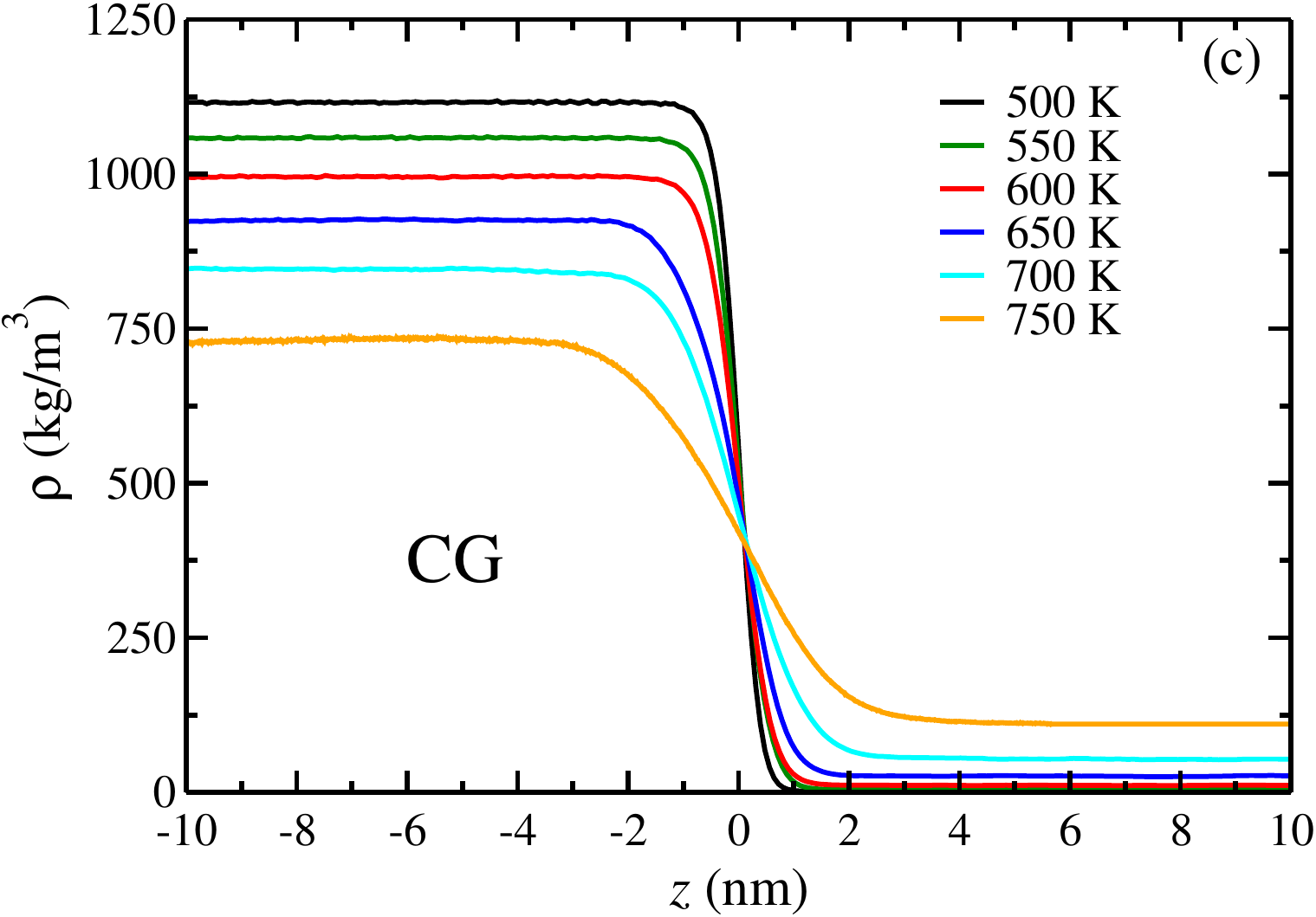}
\caption{Density profiles of HQ along one of the VL interfaces ($z$-axis) as obtained from MD simulations using the TraPPE-Hybrid-UA (a), OPLS-MOD (b), and CG (c) HQ models. The temperatures at which each density profile is obtained are represented in the legend.}
\label{figure2}
\end{figure}

The density profiles are obtained by dividing the simulation box, along the $z$-axis, in 200 slabs when simulations are carried out using both OPLS and TraPPE HQ models and in 500 slabs when the CG HQ model is used. Then, molecular density profiles are obtained along the $z$-axis, $\rho (z)$, arbitrary chosen as the direction perpendicular to the planar interfaces developed during the simulations. Molecular density profiles are obtained by assigning the position of each interaction center (atom or UA/CG group) to the corresponding slab. Finally, mass density profiles are obtained by multiplying the atom/group density profile by the corresponding assigned mass. Fig.~\ref{figure2} shows the mass density profiles obtained from the $NVT$ MD simulations using (a) the TraPPE-Hybrid-UA, (b) OPLS-MOD, and (c) CG models. We do not show the density profiles obtained using the original TraPPE-AA and OPLS models since the qualitative behavior is identical in all cases. Also, to avoid repetition, we only present one of the interfaces although the simulation box presents a V--L--V sandwich configuration with two VL interfaces.

As can be observed in Fig.~\ref{figure2}, liquid and vapor phases are presented on the left and right sides of the mass density profiles, respectively. In all cases, density profiles are obtained at $500$, $550$, $600$, $650$, $700$, and $750\,\text{K}$. When the temperature is increased, the density of the liquid phase decreases, and the density of the vapor phase increases. 

From the analysis of the density profiles, it is possible to determine the VL interfacial thickness. This property is obtained by fitting the following expression from the original mean field van der
Waals theory:~\cite{Rowlinson1982a,Rowlinson1982b}

\begin{equation}
\rho\left(z\right)=\frac{\rho_{L}+\rho_{V}}{2}-\frac{\rho_{L}-\rho_{V}}{2}\tanh\left[\frac{\alpha\left(z-z_{0}\right)}{d}\right]
\label{tanshape}
\end{equation}

\noindent
where the constant $\alpha=2\tanh^{-1}(0.8)$ is chosen so that $d$ is the 10-90
interfacial thickness and $z_{0}$ the position of the Gibbs dividing surface. The vapor and liquid coexistence densities, $\rho_V$ and $\rho_L$ respectively, are also obtained from the analysis of the density profiles as is explained in Section IV.B. $d$ and $z_{0}$ are treated as adjustable parameters and calculated by fitting density profiles to Eq.~\eqref{tanshape}. The interfacial thicknesses obtained at each temperature for each molecular model employed in this work are presented in Table \ref{table-2}.

In all cases, it can be observed how the interfacial thickness increases as the temperature increases. This is the expected behavior of the VL phase equilibria behavior of a pure system since the density of the vapor and liquid phases becomes equal at the critical temperature, and the interfacial thickness diverges to infinity at the critical temperature both phases become indistinguishable. 

\begin{table}
\centering
\caption{\label{table-2} Equilibrium liquid, $\rho_L$, and vapor, $\rho_V$, densities, vapor pressure, $P$, interfacial tension, $\gamma$, and $10-90$ interfacial thickness, $d$, obtained from $NVT$ MD simulations at different temperatures using five different models of HQ. The values in parentheses represent the error over the last two digits of the results, except for the CG results which represent the error over the last digit.
} 
\small\begin{tabular}{c c c c c c }
\hline
\hline

$T$ (K) & $\rho_L$ (kg/m$^3$) & $\rho_V$ (kg/m$^3$) & $P$ (bar) & $\gamma$ (mN/m) & $d$ (nm) \\
\hline
\multicolumn{6}{c}{TraPPE-AA} \\
\hline
500 & 1080.4(16) & 0.97(10) & 0.450(26) & 40.40(45) & 0.3989 \\
550 & 1030.3(13) & 3.25(16) & 1.571 (31) & 32.68(35) & 0.4513 \\
600 & 975.6(15) & 9.18(34) & 4.26(10) & 25.97(45) & 0.8677 \\
650 & 914.1(23) & 21.61(77) & 9.740(71) & 18.15(30) & 0.7055 \\
700 & 842.3(16) & 43.66(60) & 19.38(16) & 11.60(32) & 1.615 \\
750 & 753.9(24) & 87.43(97) & 35.26(22) & 5.74(26) & 1.595 \\
\hline
\multicolumn{6}{c}{TraPPE-Hybrid-UA} \\
\hline
500 & 1119.3(14) & 0.349(48) & 0.106(13) & 51.54(58) & 0.3407 \\
550 & 1077.67(12) & 1.18(53) & 0.476(31) & 44.13(32) & 0.4704 \\
600 & 1033.63(14) & 3.46(21) & 1.512(39) & 36.54(27) & 0.5146 \\
650 & 987.2(14) & 7.82(29) & 3.772(50) & 29.87(48) & 0.4846 \\
700 & 936.4(18) & 16.53(43) & 8.04(11) & 23.49(27) & 0.5740 \\
750 & 879.4(16) & 32.55(62) & 15.54(11) & 17.16(27) & 0.7503 \\
\hline
\multicolumn{6}{c}{OPLS} \\
\hline
500 & 1067.0(24) & 0.92(14) & 0.433(17) & 38.10(42) & 0.1917 \\
550 & 1012.6(20) & 3.38(26) & 1.527(48) & 30.97(34) & 0.5917 \\
600 & 952.1(21) & 10.06(43) & 4.432(82) & 23.31(26) & 0.6228 \\
650 & 883.4(28) & 34.90(71) & 10.64(15) & 15.60(39) & 1.292 \\
700 & 801.2(26) & 55.5(13) & 22.45(14) & 8.63(24) & 1.360 \\
750 & 686.8(28) & 120.0(14) & 41.95(22) & 3.33(26) & 2.368 \\
\hline
\multicolumn{6}{c}{OPLS-MOD} \\
\hline
500 & 1076.5(15) & 0.734(99) & 0.296(17) & 41.02(41) & 0.3779 \\
550 & 1022.9(18) & 2.93(21) & 1.201(61) & 32.17(50) & 0.4677 \\
600 & 962.9(11) & 9.05(42) & 3.902(76) & 24.30(27) & 0.9681 \\
650 & 895.6(21) & 22.38(0.63) & 9.78(10) & 16.41(38) & 0.9301 \\
700 & 814.29(28) & 49.41(91) & 20.85(18) & 10.88(34) & 1.345 \\
750 & 704.12(23) & 109.1(19) & 39.19(18) & 4.50(23) & 2.002 \\
\hline
\multicolumn{6}{c}{CG} \\
\hline
500 & 1116.18(5) & 1.40(1) & 0.54(2) & 37.35(4) & 0.3204 \\
550 & 1058.00(8) & 4.49(5) & 1.82(6) & 29.98(5) & 0.4419\\
600 & 995.10(4) & 11.41(8) & 4.76(1) & 22.22(1) & 0.5492\\
650 & 924.24(2) & 26.79(7) & 11.21(3) & 15.73(9) & 0.8545\\
700 & 844.44(4) & 54.14(3) & 21.84(3) & 9.90(3) & 1.110\\
750 & 729.21(6) & 113(9) & 39.57(5) & 4.26(5) & 1.694\\

\hline
\hline
\end{tabular}

\end{table}

\subsection{VL coexistence densities}

\begin{figure}
\centering
\includegraphics[width=0.9\columnwidth]{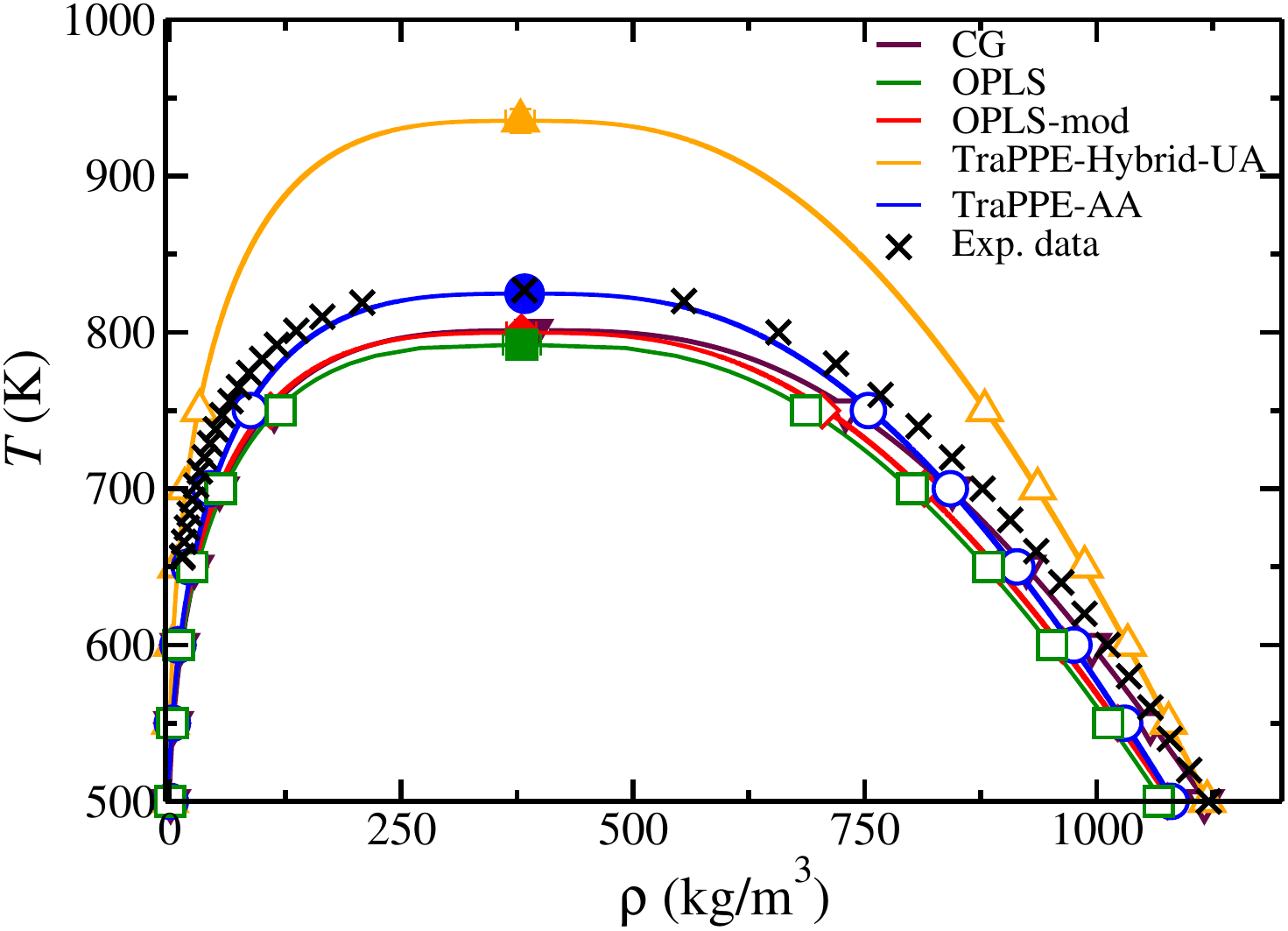}
\caption{VL coexistence densities of HQ as obtained from MD $NVT$ simulations. The five HQ models are represented as follows: TraPPE-AA (blue circles), TraPPE-Hybrid-UA (orange up triangles), OPLS (green square), OPLS-modified (red diamonds), and CG (purple down triangles). Open symbols correspond to VL coexistence density values obtained from the analysis of the density profiles. VL coexistence density curves are obtained from Eqs.~\eqref{tcscaling} and \eqref{rhocscaling}. Filled symbols correspond to the critical temperature and density estimations obtained from the fitted VL coexistence density curves. Black crosses correspond to experimental data taken from the NIST database~\cite{NIST_AM}.}
\label{figure3}
\end{figure}


VL coexistence equilibrium densities are obtained from the analysis of the density profiles. The vapor, $\rho_V$, and liquid, $\rho_L$, equilibrium bulk densities are determined by averaging the $\rho(z)$ values of the corresponding bulk phase at each temperature. Also, we only take those $\rho(z)$ values far enough from the interface to avoid any interfacial effect on the equilibrium density determination. The statistical uncertainty of the equilibrium bulk densities is estimated as the standard deviation of the mean values. VL coexistence equilibrium densities are presented in Fig.~\ref{figure3} and in Table~\ref{table-2}. As can be seen, the qualitative behavior is the same for the five models. Although the five HQ models present the same qualitative behavior, we can observe differences in the predicted densities. We also include in Fig.~\ref{figure3} experimental VL coexistence density data taken from the NIST database \cite{NIST_AM}, as well as the experimental values of $T_c$ and $\rho_c$. Here, it is important to remark that there are also two additional experimental $T_c$ values reported in the literature ($818-820\,\text{K}$)~\cite{Chleck1959a,Liessmann1995a}. As we explain in Section 3, in this work we determine $T_c$ and $\rho_c$ by applying Eqs.~\eqref{tcscaling}-\eqref{gammacscalng}. The estimated critical properties obtained from the $NVT$ MD simulations, using the five different models, are presented in Table~\ref{table3}. We observe that the VL coexistence densities, $T_c$, and $\rho_c$ values predicted by the original TraPPE-AA HQ model provide the best agreement with the experimental data available.

As we can see in Fig.~\ref{figure3} and in Table~\ref{table-2}, the VL coexistence densities predicted by both OPLS HQ models are in very good agreement between them. When we compare the OPLS results with those obtained by the TraPPE-AA HQ model and the experimental data, we notice that at high temperatures both OPLS models overestimate the coexistence vapor density. Also, it is noticeable that both OPLS models slightly underestimate the liquid coexistence density in the whole range of studied temperatures. The differences between the experimental data and OPLS simulation results become larger as the temperature is increased. As can be observed in Fig.~\ref{figure3} and Table~\ref{table-2}, the OPLS-MOD and the original OPLS models slightly underestimate the experimental $T_c$ value by $\thicksim20$ and $30\,\text{K}$, respectively although both predict very accurately the experimental $\rho_c$.

\begin{figure}
\centering
\includegraphics[width=0.80\columnwidth]{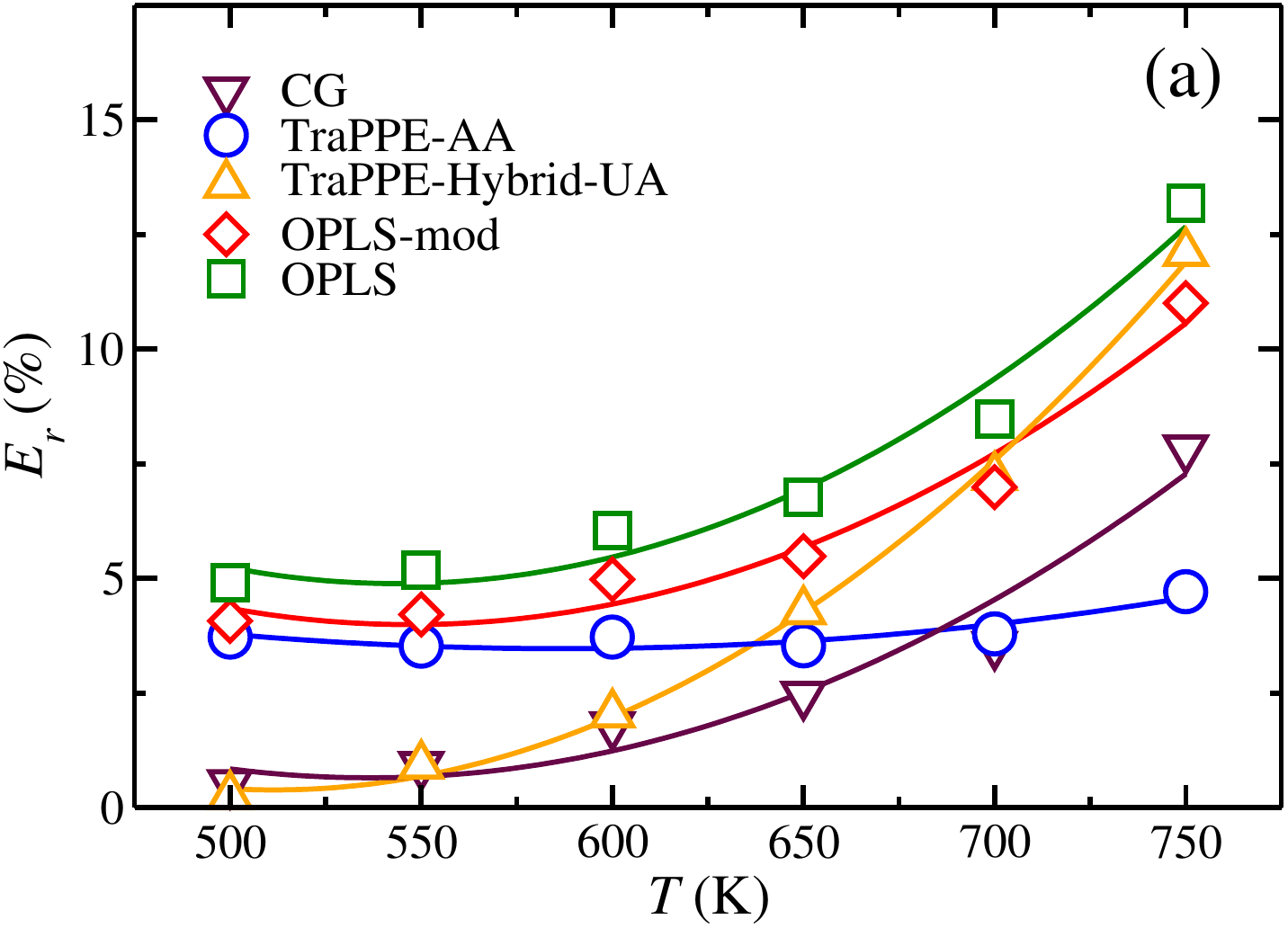}

\vspace{1cm}

\includegraphics[width=0.83\columnwidth]{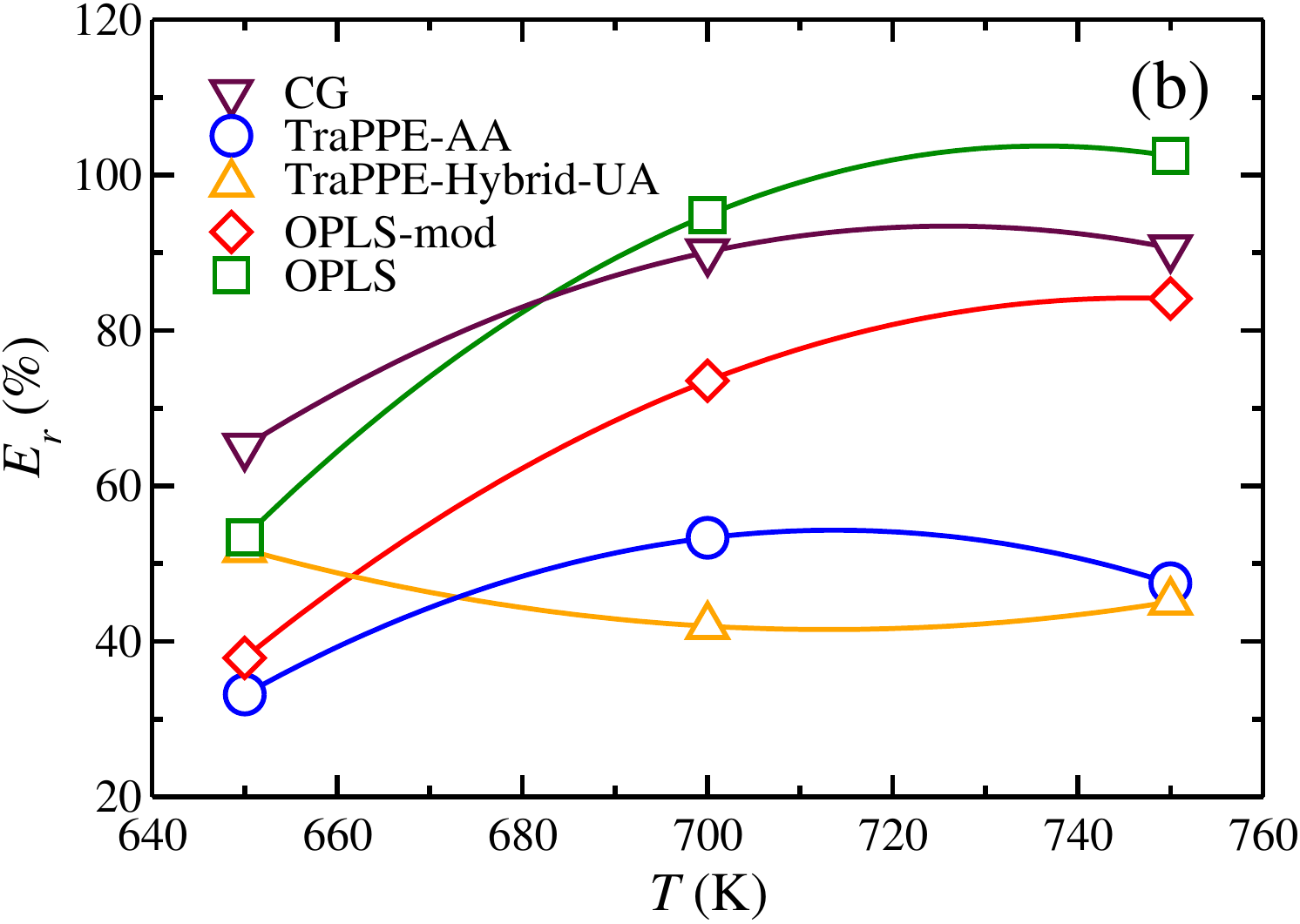}
\caption{Relative errors of the density simulation results obtained in this work by the 5 different HQ models with respect to the experimental data reported in the literature.\cite{NIST_AM} The meaning of the colors and symbols is the same as in Fig. \ref{figure2}. The top (a) figure represents the relative errors obtained for the equilibrium liquid densities and the bottom (b) figure for the equilibrium vapor densities.}
\label{Rho-error}
\end{figure}

The CG HQ model proposed in this work provides nearly identical VL coexistence results as those obtained by both OPLS models. The coexistence vapor densities obtained by the CG HQ model slightly overestimate the experimental vapor densities taken from the literature. Also, we observe that the coexistence liquid density obtained by the CG model is in excellent agreement with the experimental data at low temperatures. However, as the temperature increases, the liquid densities predicted by the CG HQ model underestimate the experimental coexistence liquid densities. The $T_c$ value predicted by the CG HQ model is similar to the $T_c$ value predicted by the OPLS-MOD model, underestimating the experimental $T_c$ value by $\thicksim20\,\text{K}$. As well as in the case of both OPLS models, the $\rho_c$ value obtained by the CG HQ model is in very good agreement with the experimental data (see Table~\ref{table-2}). 

We also discuss the results obtained by the TraPPE-Hybrid-UA HQ model proposed in this work. Unfortunately, when we compare the results obtained by this model with the predictions obtained by the rest of the models and the experimental data, the TraPPE-Hybrid-UA model underestimates the vapor coexistence density at intermediate and high temperatures and overestimates the liquid coexistence density in almost the whole range of temperatures considered in this work. As a consequence, the $T_c$ value predicted by this model overestimates the experimental $T_c$ value by $\thicksim110\,\text{K}$. It is worth mentioning that the scaling laws predict very accurately the critical conditions when the data used to fit the equations are close to the critical value. In the TraPPE-Hybrid-UA model case, the highest simulated temperature ($750\,\text{K}$) is far from the critical temperature, which means that the predicted $T_c$ value could change if higher temperature results are used to fit the scaling law equations. However, due to the poor agreement with the rest of the models, we have decided to not perform extra simulations to improve the $T_c$ predicted value using the TraPPE-Hybrid-UA model. Also, it is interesting to mention that the $\rho_c$ value predicted by this model is in good agreement with the experimental data and the results obtained by the rest of the HQ models. At this point is important to remark that the TraPPE-Hybrid-UA model is built up by mixing the TraPPE-UA parameters of the benzene molecule with the TraPPE-AA parameters of the aromatic-OH group from the original TraPPE-AA HQ molecule. This combination has been made in order to improve the speed of the HQ original TraPPE-AA model but it is important to notice that the parameters from both TraPPE approaches have not been developed to be mixed. This could explain why the results trend of the TraPPE-Hybrid-UA HQ model is different from the other models. A possible solution to improve the predictions at high temperatures would be to re-parametrize the local charges placed at each chemical group~\cite{Comesana2018a,Bhatnagar2013a}. Some of the authors of this work used this procedure to recalculate the partial charges located at each chemical group of the OPLS-MOD HQ model~\cite{Comesana2018a}. In particular, it would be interesting to recalculate the partial charges located at the --OH groups of the new TraPPE-Hybrid-UA model which are probably responsible for the high predictive critical temperature due to hydrogen-bond formation. However, the optimization of the parameter models is out of the scope of this work.

Finally, we calculate the relative errors between the simulation equilibrium density results and the experimental data~\cite{NIST_AM} by using the following equation:

\begin{equation}
    E_r(\%)=\frac{P_{exp}-P_{sim}}{P_{exp}}\times100
    \label{Rel_error}
\end{equation}

\noindent where $P_{exp}$ and $P_{sim}$ stand for the property under inspection obtained from experiments and from simulations respectively. This procedure is applied to the results obtained by the 5 different HQ models used in this work. Figure \ref{Rho-error} shows the density relative errors calculated in the range of temperatures where experimental data is available. We also plotted a fitted curve of the results obtained from Eq.~\eqref{Rel_error} to show the trend of the relative errors in the experimental range of temperature available. From Figure \ref{Rho-error}a, we find that for liquid densities, from low to intermediate temperatures ($500-650\,\text{K}$), the best agreement with the experimental data is obtained by the TraPPE-Hybrid-UA and the CG model, while from intermediate to high temperatures ($650-750\,\text{K}$), the best agreement is obtained by the CG and the TraPPE-AA models. From Figure \ref{Rho-error}b it is possible to analyze the agreement between experimental data and simulation results for the equilibrium vapor densities case. Unfortunately, there are just three temperatures at which vapor densities can be compared directly (650, 700, and $700\,\text{K}$), although the general trend can be intuited. In the range of temperatures considered, both TraPPE models provide the best agreement with the experimental data, although the TraPPE-AA model is expected to provide better results at lower temperatures if we take into account the trend of the relative error fitting curves.

\begin{table}[h!]
        \caption{Experimental and predicted critical properties of HQ with different Force
        Fields models}
        \begin{center}
                \begin{tabular}{ccccc}
                        Models & $\rho_{c}$ (kg/m$^{3}$) & $T_{c}^{a}$ (K)& $T_{c}^{b}$ (K) & $P_{c}$ (bar)\\
                        \hline
Experimental(1)\cite{Liessmann1995a} & - & 820 & - & {67.5} \\
Experimental(2)\cite{NIST_AM} & 383 & 827.8 & - & {127.4*} \\
{Experimental(3)\cite{Gmehling1983a}} & - & {818} & - & {61.8 } \\
OPLS & 380(20) & 792.2(61) & 790(11) & 68.5(38) \\
OPLS-MOD & 380(16) & 799.8(50) & 802(11) & 74.0(19)\\
TraPPE-AA & 383(15) & 824.8(54) & 817(13) & 76.7(34)\\
TraPPE-Hybrid-UA & 379(16) &953.3(76)  & 929(14) & 99.6(56) \\
CG&393.36(56)  & 801.10(18) & 803.2(16) &68.1(13) \\
                        \hline
                \end{tabular}
        \end{center}
        \scriptsize 
        Critical densities ($\rho_{c}$), temperatures ($T_{c}^{a}$ and $T_{c}^{b}$), and pressure ($P_{c}$) of HQ, as obtained from the five different models used in this work. $\rho_{c}$ and $T_{c}^{a}$ are obtained from the analysis
        of the coexistence densities using  Eqs.~(\ref{tcscaling}) and (\ref{rhocscaling}), $T_{c}^{b}$ is obtained from the analysis of the interfacial tension results using Eq.~(\ref{gammacscalng}), and, finally, $P_{c}$ is obtained from the analysis of the vapor pressure results and using Eq.~(\ref{pcscaling}). 
        {*Experimental critical vapor pressure is obtained from the analysis of the experimental vapor pressure data and using Eq.~(\ref{pcscaling}).}
        \label{table3}
\end{table}

\subsection{Vapor pressure}

\begin{figure}
\centering
\includegraphics[width=0.9\columnwidth]{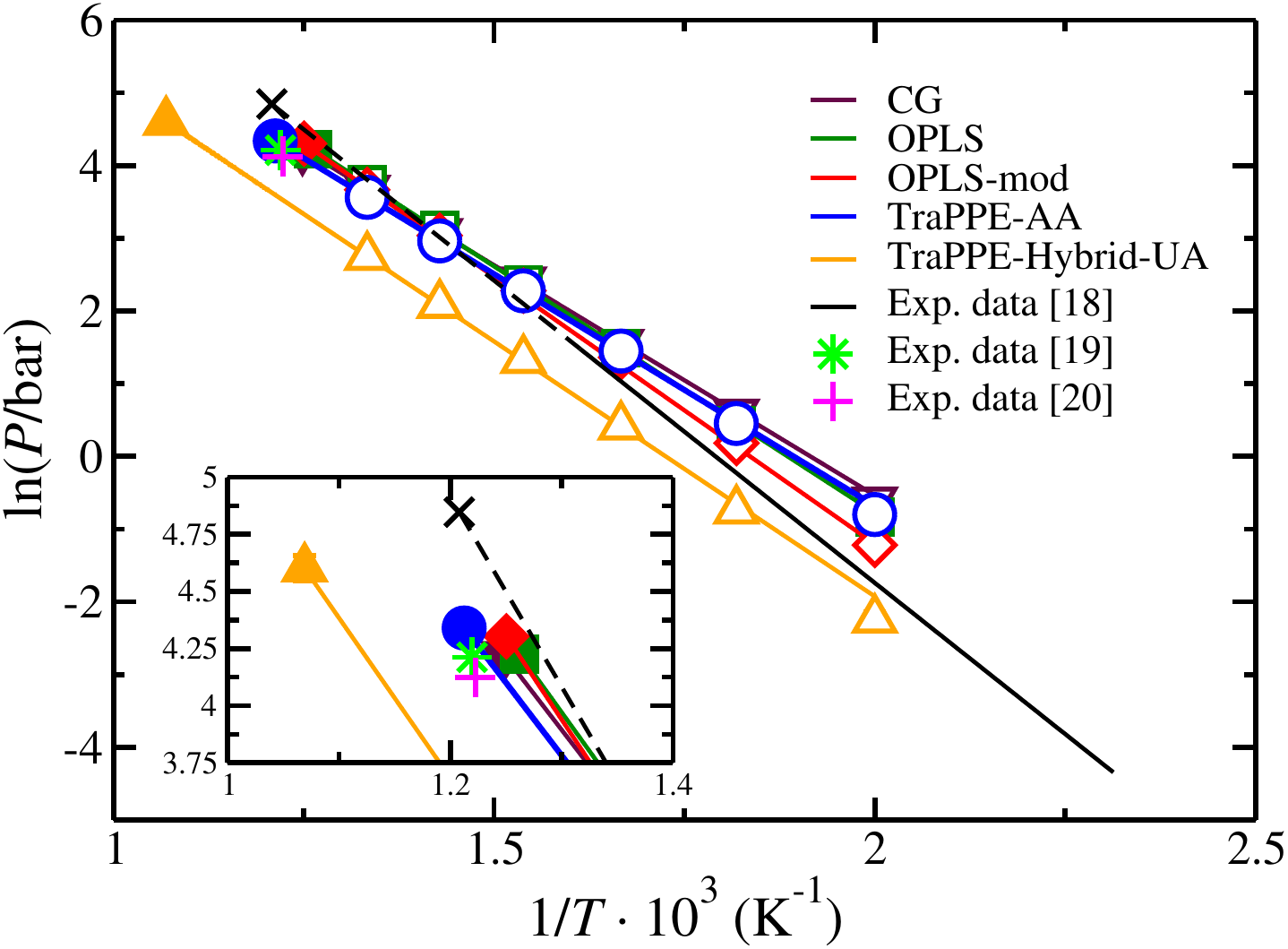}
\caption{Natural logarithm of the vapor pressure vs $1/T$ of HQ as obtained from MD $NVT$ simulations. The meaning of the colors and symbols is the same as in Fig.~\ref{figure3}. Vapor pressure lines are obtained from Eq.~\eqref{pcscaling}. Filled symbols correspond to the critical pressure estimations obtained from Eq.~\eqref{pcscaling} using the $T_{c}^a$ value presented in Fig.~\ref{figure3} and in Table \ref{table3}. The black curve corresponds to experimental data taken from the literature~\cite{NIST_AM} and the black cross is the estimated critical pressure obtained by fitting the experimental data~\cite{NIST_AM} using Eq.~\eqref{pcscaling}. The pink plus and the green star correspond to experimental data taken from the literature.\cite{Liessmann1995a,Gmehling1983a} We also represent a zoom of the critical conditions as inset.}
\label{figure4}
\end{figure}


As we have explained previously, we determine the equilibrium vapor pressure as the $P_{zz}$ component of the pressure tensor since both phases are in contact via a planar interface along the $z$-axis direction. We present the results obtained in this work in Fig.~\ref{figure4} and Table~\ref{table-2}. We also include experimental data taken from the literature \cite{NIST_AM} in Fig.~\ref{figure4}. We also determine the critical pressure value, $P_c$, predicted by the five models using Eq.~\eqref{pcscaling} and the $T_c$ value obtained from the fitting of the VL coexistence densities, $T_{c}^a$. The predicted $P_c$ values are presented in Fig.~\ref{figure4} and Table~\ref{table3}.



\begin{figure}
\centering
\includegraphics[width=0.83\columnwidth]{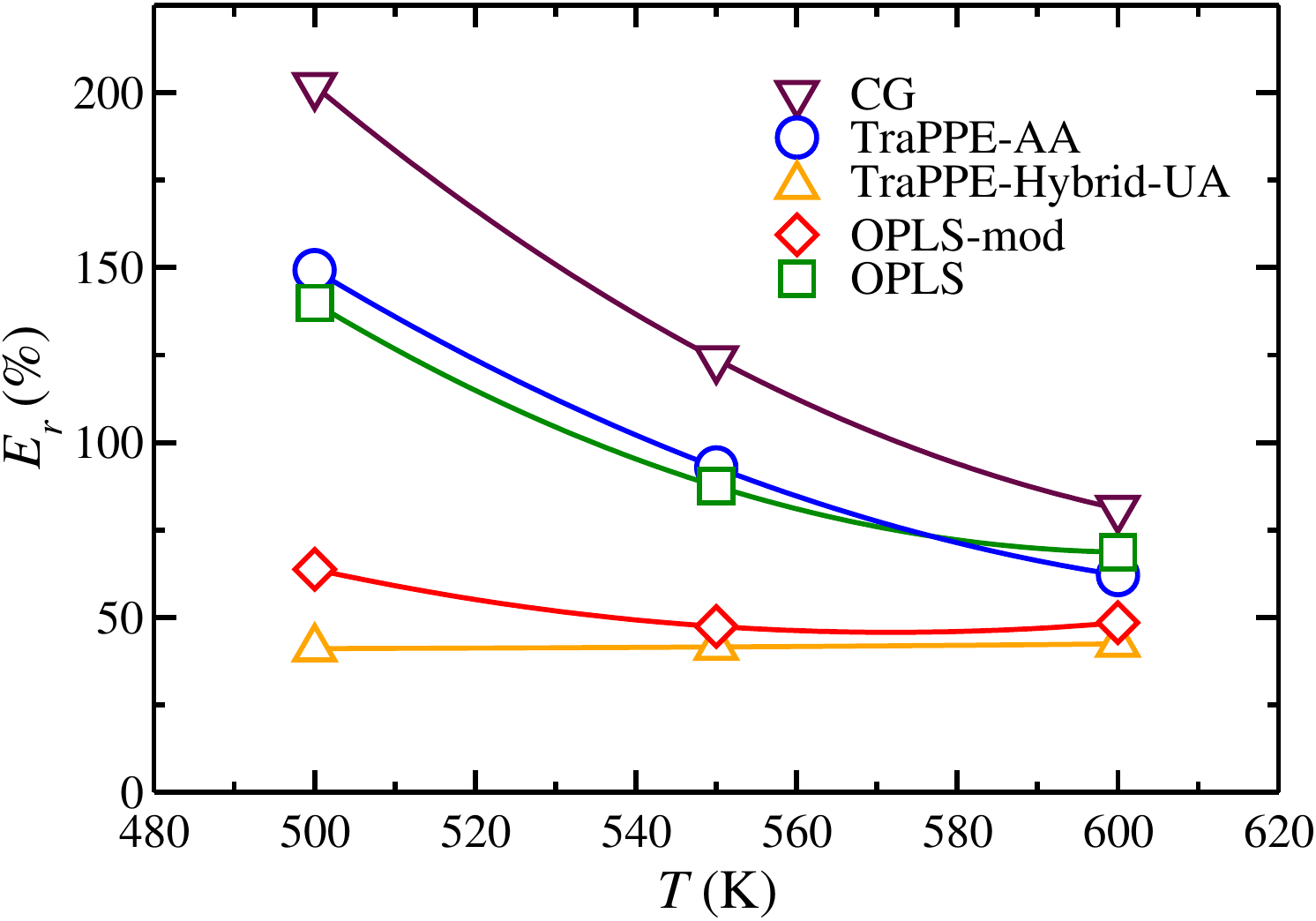}
\caption{Relative errors of the vapor pressure simulation results obtained in this work by the 5 different HQ models with respect to the experimental data reported in the literature~\cite{NIST_AM}.}
\label{P-error}
\end{figure}

As we can see, at low temperatures (from 500 to $620\,\text{K}$), the vapor pressure values predicted by the five HQ molecular models and the experimental data are in good agreement. Fig.~\ref{figure4} shows that the different HQ models follow the same trend as in Fig.~\ref{figure3}. The TraPPE-AA, CG, and both OPLS HQ models provide nearly the same results. Finally, and following the same trend as in  Fig.~\ref{figure3}, the predictions obtained by the TraPPE-Hybrid-UA HQ model underestimate the predictions obtained by the rest of the HQ models. In order to quantify which model provides a better description of the HQ vapor pressure, we calculate the relative errors between simulation results and the scarce experimental data available in the literature. As we can see in Figure \ref{P-error}, the best agreement is obtained by the TraPPE-Hybrid-UA HQ model, closely followed by the OPLS-MOD HQ model. Unfortunately, the range of temperature at which vapor pressure experimental data is available in the literature is narrow (from 500 to $600\,\text{K}$)~\cite{NIST_AM}. If we take into account the trend shown by each model in Figure \ref{P-error}, it seems that agreement between experimental data and simulation results is improved at higher temperatures. Unfortunately, there is no vapor pressure experimental data reported beyond $600\,\text{K}$. However, in order to estimate which model provides a better description of the vapor pressure at higher temperatures, we compare the predicted critical vapor pressure by using Eq.~\eqref{pcscaling} with the experimental values reported in the literature. According to the experimental data reported by Liessmann \emph{et al.}~\cite{Liessmann1995a} and Gmehling \emph{et al.}~\cite{Gmehling1983a}, the best two models for predicting the experimental $P_c$ are the CG and the original OPLS models. The rest of the models overestimate the $P_c$  value, being the TraPPE-Hybrid-UA model the one that further overestimates it. We have also estimated the experimental critical vapor pressure by using equation \eqref{pcscaling} and the reported experimental data~\cite{NIST_AM}. The estimated critical vapor pressure is represented in Fig. \ref{figure4} and in Table \ref{table-2}. The estimated experimental $P_c$ value overestimates by far the experimental values reported in the literature, as well as the predictions obtained from molecular dynamic simulations. As can be seen, all the models underestimate the predicted critical vapor pressure value. The relative errors obtained by each model are 22, 40, 42, 46, and $46\%$ for the TraPPE-Hybrid-UA, TraPPE-AA, OPLS-MOD, CG, and OPLS, respectively. From this analysis, we can conclude that the TraPPE-Hybrid-UA model and the OPLS-MOD models provide the best agreement with the estimated and reported experimental data in the whole range of temperatures considered in this work, however, although the original OPLs and TraPPE-AA models present larger deviations of the vapor pressure at low temperatures than the OPLS-MOD and TraPPE-Hybrid-UA models, they provide better agreement with the experimental critical $P_C$ value predicted in the literature. This conclusion invites us to think the critical vapor pressure can not be estimated from the fitting by using the experimental vapor pressure at low temperatures and equation \eqref{pcscaling} and more experimental point at higher temperatures are required.

\subsection{Interfacial tension}

\begin{figure}
\centering
\includegraphics[width=0.9\columnwidth]{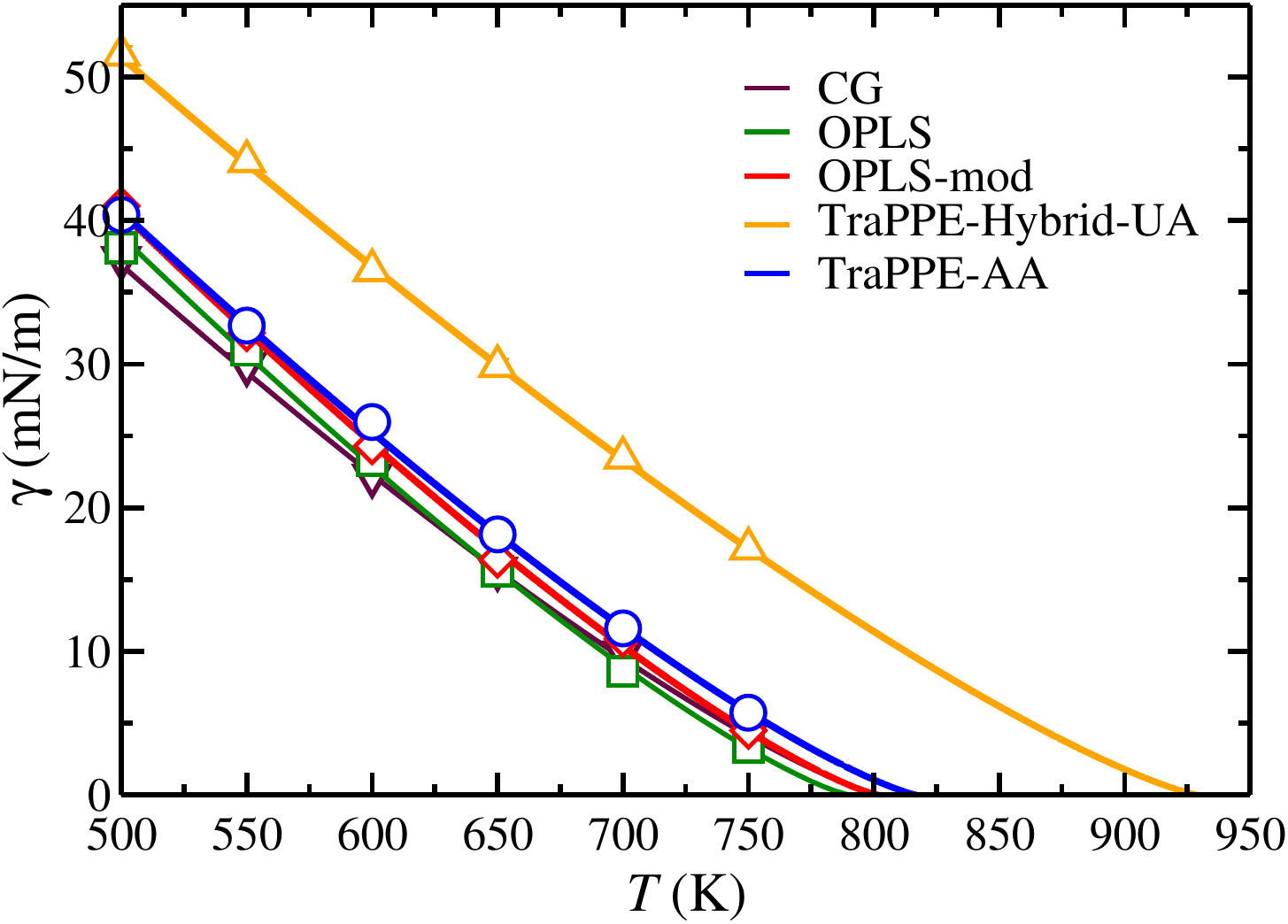}
\caption{VL interfacial tension of HQ as obtained from MD $NVT$ simulations. The meaning of the colors and symbols is the same as in Figs.~\ref{figure3} and \ref{figure4}. The curves are obtained from Eq.~\eqref{gammacscalng}.}
\label{figure5}
\end{figure}

Finally, we determine the VL interfacial tension, $\gamma$, from the components of the pressure tensor using Eq.~\eqref{iftmd}. The interfacial tension property is one of the most sensitive to the molecular models and simulation details. In this work, simulation details are similar in all cases but we employ five different molecular models. Interfacial tension results obtained in this work are presented in Fig.~\ref{figure5} and Table \ref{table-2}. As we can see, agreement between the results obtained by the different models exhibits a similar behavior as those observed for the VL coexistence densities and vapor pressure. The predictions obtained using the TraPPE-AA, CG, and both OPLS models are in very good agreement, while the results obtained by the TraPPE-Hybrid-UA model clearly overestimate the results predicted by the rest of the models. Unfortunately, as far as the authors know, there is no experimental data available in the literature and we cannot compare our predictions with experimental results.

Here it is interesting to remark on the results obtained by the CG HQ model. This model provides results in very good agreement with the TraPPE-AA and both OPLS models, which are the most realistic and computationally expensive HQ models. As it has been explained in Section II.C, the CG HQ model considers the hydroquinone molecule as a planar one formed from four identical segments without local charges. It is a very simple rigid model since it is not necessary to take into account intramolecular bonded interactions, which allows the use of larger simulation time steps. Also, there are no charges and it is not necessary to calculate non-bonded coulombic interactions. As a consequence of the simplicity of the CG HQ model, simulations performed with this model are cheaper than those performed with the TraPPE and OPLS models, where more charged interactive sites are necessary to describe the HQ molecule and it is also necessary to use lower simulation time steps to account for the flexibility of the models. It is important to remark that in all cases the thickness of the liquid phase is big enough to avoid any size effect on the surface tension determination~\cite{Ivanova2023a}. In particular, the liquid phase thickness of simulations carried out with both TraPPE and OPLS models is $\sim 20$ times the HQ molecular size. In the case of the CG model, the liquid thickness is even bigger and it is $\sim 30$ times the CG molecular size.

\section{Conclusions}

In this work, we determine the VL equilibria and interfacial properties of the HQ system using five different molecular models. The first two models are based on the TraPPE force fields~\cite{Rai2007a,Rai2013a,Wick2000a}, modeling the HQ molecule following an all-atom (TraPPE-AA) and a hybrid united-atom (TraPPE-Hybrid-UA) approaches. The next two models are based on the OPLS force fields. Both models are based on an all-atom approach where the first one is the original OPLS model~\cite{Jorgensen1993a,Jorgensen1996a}, and the second one is a modified version where the local charges located at each atom are reparametrized (OPLS-MOD)~\cite{Comesana2018a}. These four models provide a realistic representation of the HQ molecule since they take into account, not only the functional group of the HQ molecule but also the flexibility and internal degrees of freedom. Finally, following the CG approach, the HQ molecule is modeled as a rigid and planar molecule formed by four identical tangent spheres. These segments interact through the Mie potential and there are no local charges placed on them. 

Molecular dynamics simulations are carried out in the $NVT$ ensemble using the direct coexistence technique. Following this approach, both phases in equilibrium (vapor and liquid) coexist, sharing an interface, in the same simulation box. This technique allows us to study not only the phase equilibria properties but also the interfacial ones since a stable VL interface exists in the simulations. In particular, we determine the density profiles, VL coexistence densities, vapor pressures, interfacial thicknesses, and interfacial tension. These properties are studied from 500 to $750\,\text{K}$ using the five molecular models employed in this work. Finally, we also determine the critical temperature, density, and pressure predicted by the five models.

The VL coexistence densities predicted by the five models are in very good agreement except for the predictions obtained by the TraPPE-Hybrid-UA model. This model underestimates/overestimates the vapor/liquid coexistence densities which result in a wider VL phase envelope. We conclude that the CG, both OPLS and the TraPPE-AA models can be used to describe very accurately the experimental VL phase behaviour of the HQ molecule, although the TraPPE-AA model provides the most accurate description of the VL coexistence densities. Also, among the five models studied in this work, the critical temperature obtained from the original TraPPE-AA model shows the best agreement with the experimental critical temperature values, followed by the CG, OPLS-MOD, OPLS, and TraPPE-Hybrid-UA models in this order.


We also compare the estimated vapor pressure values with experimental data taken from the literature~\cite{NIST_AM}. Unfortunately, the experimental vapor pressure of the HQ system has been determined at low temperatures (from 500 to $620\,\text{K}$). The vapor pressure predicted by the five models is in good agreement with the experimental data. Again, the original TraPPE-AA model seems to provide the best agreement with the experimental values, although this model slightly overestimates the experimental results. The CG and both OPLS models also provide a good agreement with the experimental values and, following the same trend as the TraPPE-AA model, they slightly overestimate the experimental results. Finally, the TraPPE-Hybrid-UA model also provides a good agreement with experiments, but in this case, the predictions obtained by this model underestimate the experimental data. Since there is no experimental data at higher temperatures, we cannot assess which model is superior over the others for predicting vapor pressure values. However, due to the trend of the experimental value and the good agreement between the critical temperature predicted by the TraPPE-AA model and the experimental data, we suggest the TraPPE-AA as the best option for an accurate description of the VL equilibria behavior of the HQ pure system. At this point, it is relevant to remark that the pressure values predicted by the CG model are in excellent agreement with those provided by the TraPPE-AA model. This is important since the CG model is much simpler and cheaper to simulate than the TraPPE and OPLS models. As a consequence, simulations carried out using the CG model are about five times faster than those using more realistic models such as TraPPE and OPLS models.

Finally, we determine the interfacial tension of HQ as a function of temperature. This property is extremely sensitive to molecular model details. Due to the scarcity of experimental data for HQ, we cannot compare our findings with experimental results taken from the literature. In this case, we compare the predictions obtained from the different models. The trend of the results is similar to those for the rest of the properties. The agreement between the interfacial tension results obtained by the TraPPE-AA, CG, and both OPLS models is excellent in the whole range of studied temperatures. The predictions obtained by the TraPPE-Hybrid-UA model clearly overestimated the results obtained by the other four models. This is an expected result since this model overestimates also the critical temperature and underestimates the vapor pressure. Although some of the molecular HQ models employed in this work have already been reported in the literature, this is the first time that they have been used to describe the VL equilibria and interfacial properties of HQ. We expect that our results provide a better comprehension of its VL phase diagram and interfacial properties.

\section*{CRediT authorship contribution statement}

\textbf{Miguel J. Torrejón}: Investigation, Writing – review \& editing. \textbf{Brais Rodríguez García}: Investigation, Writing – review \& editing. \textbf{Jesús Algaba}: Conceptualization, Methodology, Supervision, Writing - original draft, Writing – review \& editing. \textbf{José Manuel Olmos}: Investigation, Writing – review \& editing. \textbf{Martín Pérez-Rodríguez}: Investigation, Methodology, Writing – review \& editing. \textbf{José Manuel Míguez}: Investigation, Writing – review \& editing. \textbf{Andrés Mejía}: Investigation, Methodology, Writing – review \& editing. \textbf{Manuel M. Piñeiro}: Conceptualization, Investigation, Methodology, Project administration, Writing – review \& editing. \textbf{Felipe J. Blas}: Conceptualization, Investigation, Methodology, Supervision, Project administration, Writing – review \& editing.

\section*{Declaration of competing interest}

The authors declare that they have no known competing financial interests or personal relationships that could have appeared to influence the work reported in this paper.

\section*{Acknowledgements}

This work was funded by Ministerio de Ciencia e Innovaci\'on (Grant No.~PID2021-125081NB-I00 and No.~PID2024-158030NB-I00) and Universidad de Huelva (P.O.~FEDER~EPIT1282023), both co-financed by EU FEDER funds. MJT also acknowledges the research contract (Ref.~01/2022/38143) of Programa Investigo (Plan de Recuperaci\'on, Transformaci\'on y Resiliencia, Fondos NextGeneration EU) from Junta de Andaluc\'{\i}a (HU/INV/0004/2022). MPR acknowledges grant Ref. CNS2022-135881 financed by MCIN/AEI/10.13039/501100011033 and NextGenerationEU/PRTR. We greatly acknowledge RES resources at Picasso provided by The Supercomputing and Bioinnovation Center of the University of Málaga to FI-2024-1-0017.

\section*{Supplementary material}

The parameters of all the molecular models employed in this work can be found online.

\section*{Data availability}

No data was used for the research described in the article.




 \bibliographystyle{elsarticle-num} 
\bibliography{masterbib}





\end{document}